\documentstyle[12pt,fleqn]{article}
\newcommand{\talkauthors}[1]{
{\small {\bf  \enskip \enskip #1}}}
\newcommand{\talktitle}[1]{{\small {\bf #1}}}
\newcommand{\address}[1]{{\small {\it #1}}}

\def\pbar{\overline{p}}

\begin{document}
\vskip 4mm
\begin{flushright}
FNT/AE 93-25\\
FIAN/TD-15/93\\
July  1993
\end{flushright}
\begin{center}

\vskip 20mm

\talktitle{CUMULANT TO FACTORIAL MOMENT RATIO\\ AND MULTIPLICITY DATA}
\vskip 15mm
\talktitle{I.M.Dremin\footnote{Work
partially supported by the Russian Fund for Fundamental
Research, grant 93-02-3815 and by the NATO grant CRG 930025.}}\\
\address{P.N.Lebedev Institute of Physics,\\
Academy of Sciences of the Russia, Moscow, Russia}
\vskip 7mm
\talkauthors{
V.Arena, G.Boca, G.Gianini,\\
S.Malvezzi, M.Merlo, S.P.Ratti, C.Riccardi,\\
G.Salvadori, L.Viola, P.Vitulo
}\\
\address{
Department of Nuclear and Theoretical Physics \\
University of Pavia and Sezione INFN, Pavia, Italy
}
\vskip 11mm
\end{center}

\begin{abstract}
The ratio of cumulant to factorial  moments of experimental multiplicity
distributions has been calculated for $e^{+}e^{-}$ and $hh$ interactions
in a wide range of energies. As a function of the rank it
exhibits an initial steep decrease and a series of oscillations around zero.
Those features cannot be reproduced by the Negative Binomial Distribution.
A comparable behaviour is instead predicted in high-energy perturbative QCD.
The presence of a qualitatively similar behaviour
for different processes and in wide energy intervals suggests speaking
of an approximate scaling of the cumulant to factorial moment ratio.
\end{abstract}

\newpage
\section{Introduction}

In this paper
we show that the ratios of cumulant to factorial moments
of the multiplicity distributions measured in $e^{+}e^{-}$-collisions
(in the energy range from 22 to 91 GeV) and those given in $hh$-collisions
(from about 24 to 900 GeV) exhibit,
as a function of the rank, a peculiar behaviour
which reproduces surprisingly well the
qualitative features predicted by
 QCD calculations \cite{Dremin:PLB},\cite{Dremin:1}
for hard processes at asymptotic energies.

 The set of moments of the multiplicity distribution for
particles produced in
high energy collisions is as representative as the distribution itself. We
shall use the normalized factorial moments $F_q$ defined as
\begin{equation}
F_{q}= \Sigma_{n=1}^{\infty} n(n-1)...(n-q+1)P_{n}/\langle n \rangle^{q}
\end{equation}
(where $P_{n}$ is the probability of $n$-particle
events, $\langle n\rangle=\Sigma n\,P_{n}$ is
the mean multiplicity) and the
cumulant moments $K_q$ related to factorial moments
by
\begin{equation}
F_q= \Sigma_{m=0}^{q-1} C_{q-1}^{m}K_{q-m}F_m
\end{equation}
with $F_0=F_1 =K_1 =1, K_0 =0$ and $C_{q}^{m}=q!/m!(q-m)!$.
For the Poisson distribution, factorial moments are identically equal to 1
and all cumulants vanish for $q>1$ (but $K_1 =1$).
Experimental distributions are
usually much wider and moments increase with their rank quite fast.
For this reason it is useful to consider the ratio
\begin{equation}
H_q =K_q /F_q
\end{equation}
whose asymptotic behaviour must be flatter.

The minimum of that ratio has been predicted at $q \approx $ 5
in next-to-next-to leading
perturbative gluodynamics
\cite{Dremin:PLB} and later it has been shown \cite{Dremin:1}
that this ratio should reveal a sequence of negative minima and positive
maxima,
as the ever higher order solution of the non-linear
equation for the
generating functions of multiplicity distributions in gluodynamics.
The role of quark degrees of freedom has been shown \cite{Dremin:2},
\cite{Dremin:3} to be not very important for the qualitative features.
They persist in the exact solution of the QCD equations in case of fixed
coupling constant as well [4].

We point out that neither a lower order QCD approximation such as Double
Logarithmic Approximation (DLA) nor Negative Binomial Distribution (NBD)
favour such a peculiar behaviour \cite{Dremin:3},\cite{Dremin:4}.
On the contrary both of them imply that $H_{q}$ is a decreasing
positive function tending at large $q$ asymptotically to zero
as a power-like function. Such decrease is monotonous both in NBD and in DLA
QCD; however, some additional wiggles could be imposed on it if
preasymptotic terms are added to the asymptotic DLA formulae \cite{Dremin:4}
(this effect is related to the narrowing of multiplicity
in higher order QCD noticed in \cite{Dokshitzer}).
Thus the existence of the minima, their positions and the value of $H_q$
at those points indicate the importance of non-leading terms in
the QCD cascade development.

Let us note that the sets of numbers $P_n, F_q, K_q, H_q$ are, formally
speaking, equivalent. In practice, however, the patterns reflecting different
dynamics could be easier revealed by plotting the most suitable of them on the
corresponding scale. One knows, e.g., that the tail of $P_n$ is better
reproduced
on log-scale while its behaviour near maximum is clearly seen on linear scale.
Even more sensitive to that tail is the behaviour of factorial
and cumulant moments at high ranks.
Nevertheless there is no clear qualitative distinction between the ever
increasing factorial moments in higher order perturbative QCD results and
NBD with properly chosen parameters even though the rate of increase
and numerical values differ \cite{Dremin:2}.
The cumulant moments are much more sensitive to the tiny details of
distributions and reveal qualitative difference between QCD and NBD
predictions. Their fast increase with the rank is somewhat damped
if one uses their ratio to factorial moments. That is why it
seems to be much more
convenient to re-map the multiplicity distributions to the ratio $H_q$.
This ratio first appeared [1] quite naturally as a solution of the
equation for the generating function in gluodynamics.
It can be easily calculated from experimental data on multiplicity
distribution $P_n$ by using (1), (2).
Therefore, we choose the ratio of cumulant to factorial moments $H_q$ as a more
sensitive measure of multiplicity distributions and show its advantages in
what follows.

In this paper the values of the ratios $K_q/F_q$ for
multiplicity data from inelastic
collisions covering $e^{+}e^{-}$ annihilation and  hadron-hadron interaction
at various energies are analyzed.

\section{Experimental Results}

A large representative sample of multiplicity data [6-16]
has been collected from the wide available literature:
due to the precision
required to see the investigated features, the low statistics
samples, yielding large error bars, have been disregarded and
only papers reporting a detailed separation between elastic and inelastic
data for low multiplicities have been taken into account.
A list of the considered samples is given in Table I.
We do not attempt here the quantitative comparison with QCD predictions
but focus our attention on the main qualitative features.
For all the considered experimental multiplicity distributions, the
ratio $K_{q}/F_{q}$ has been computed up to the 16-th order.
Comparisons with the NBD predictions have been done
using the following relation obtained in ref. \cite{Dremin:4} :
\begin{equation}
H_{q}^{NBD}=\frac{\Gamma(q)\Gamma(k+1)}{\Gamma(k+q)}\;\;.
\end{equation}

Though the involved orders of magnitude of $H_{q}$ are interaction and
energy dependent, the multiplicity distributions of both hadron-hadron
collisions and $e^{+}e^{-}$-annihilations give a qualitatively similar
behaviour.

As a first example one may consider the outcomes of the
$e^{+}e^{-}$ 91 GeV DELPHI multiplicity distribution \cite{DELPHI},
plotted in fig.1.
Due to the different  order of magnitude involved here, the
low-$q$ and high-$q$ regions are plotted using different scales.
The polyline in the upper right box displays, up to the fourth order,
the abrupt descent of $H_{q}$ (the scale is logarithmic).
In the main reference frame two negative minima,
(at $q\approx$ 5 and 12 )  and two positive maxima
(at $q\approx$ 8 and 15 ) are shown.

The $H_{q}$ value at the first minimum is
$H_{q_{1}}$=$-7.6\pm1.0\times 10^{-4}$.
The NBD predictions are given by the
dashed curve shown in fig. 1: the value used for the NBD parameter $k$
has been taken from ref. \cite{DELPHI} ($k^{-1}$=0.0411$\pm$0.0012).

Fig. 2 refers to the $\overline{p}p$ 546 GeV interaction outcomes
from the UA5 collaboration \cite{UA5:546}. The
main difference with respect to the previous case is the order
of magnitude of minima and maxima, their absolute values are
more than ten times
larger; for instance the value at the first minimum
is $H_{q1}$=$-8.3\pm2.7\times 10^{-3}$.
Nonetheless, the same main characteristics
found for  the $e^{+}e^{-}$ annihilation, i.e. the abrupt
descent and the subsequent oscillations can be observed (here minima are
at $q\approx$ 6 and 12, while maxima are at $q\approx$ 9 and 15).
Again superimposed  (dashed curve) are the NBD yields, here corresponding
to the $k$ parameter given in reference \cite{UA5:546}
($k$=3.69$\pm$0.09).

The behaviour of the $H_{q}$ ratio vs. $q$ for the other
$e^{+}e^{-}$ and hadron-hadron experiments examined
is reported in fig.s 3 and 4 respectively.
In both the $hh$ and  $e^{+}e^{-}$ outcomes, a steep descent,
taking place at the lower moment orders and showing positive
values ($K_{q}$ positive), is followed by a negative minimum
(for $q=$4, 5 or 6), then a sequence of oscillations begins.
In all the cases at least two negative minima are shown.
In all the outcomes with the exception of fig. 3b, refering to the
HRS collaboration data, the oscillations are slow and
the distance between two subsequent minima is usually 5 to 8 ranks of moment.
The close but not equal behaviour of the four LEP experiments
(fig.s 3e-3h) indicates the extreme sensitivity of the $H_{q}$
to slight differences among multiplicity distributions.
Though the features at different energies
and for $hh$ and  $e^{+}e^{-}$ interactions
are qualitatively very similar, the order of magnitude
of the oscillations differ of about a factor ten, being larger
for the hadronic interactions.

\section{Discussion and Conclusions}

We performed this investigation
prompted by the QCD prediction \cite{Dremin:PLB}
of the specific behaviour of the ratio $H_{q}$
of cumulant to factorial
moments for multiplicity distributions, with a minimum at a well
defined position which is uniquely
determined by the QCD anomalous dimension [1]. Early experimental outcomes,
though not directly comparable with theory,
raised the problem of the role in perturbative equations
of higher order terms. In ref. \cite{Dremin:1} it was proved that they
could actually give rise to an oscillating behaviour.
This peculiarity only appears in higher order approximations of QCD
and is determined solely by the gauge theory parameters
(the value of the coupling constant and the number of active
flavours) and by the kernels of the equation for the generating functions.
It has been confirmed by the exact solution of QCD equations
for fixed coupling constant \cite{Dremin:3}.

We could expect such qualitative features to be appropriate for
$e^{+}e^{-}$-annihilation since the above prediction is valid for
hard processes at high energies. Our expectations have been
confirmed. It is a surprise to observe similar
features in soft $hh$-collisions which are out of the scope of
theoretical approximations of perturbative QCD for hard processes.
We do not speculate whether this
indicates the similarity of $s$- and $t$-channel cascading in
both types of events and/or the early onset of the perturbative
QCD regime but, at the same time, we are not able to conceal our
thoughts along these lines.
The quantitative difference between $e^{+}e^{-}$ and $hh$
distributions, however, should be studied more carefully. Preliminary
analysis of $hA$-data (which we do not discuss here) implies
new possibilities too.

Thus we must state that it is an initial stage of approach to
the whole problem but
the data presented here seem to suggest the following remarks:
the proposed ratio $H_{q}$ is a sensitive measure of multiplicity
distributions;
besides the observed minima in $e^{+}e^{-}$ and $hh$
processes it reveals some other peculiarities at higher $q$ which
should be studied and helps distinguishing clearly various distributions
which are hardly separated by considering factorial moments only.
In particular  NBD predictions for $H_{q}$ are found to disagree from
experimental data: to some degree such a failure was expected in
full-phase space inclusive data; to the comparison
of the $H_{q}$ behaviour with NBD predictions in restricted phase-space
regions and fixed jet multiplicity samples a dedicated
work will be devoted \cite{NBD:Are:93}.

Furthermore the qualitative behaviour of $H_q$
is conserved for different processes
in wide energy intervals so that one can think of an
approximate scaling of $H_q$;
nonetheless quantitative differences in the size of minima are observed
and the problem of asymptotics should be considered in more details.
Experimental data on $H_{q}$ in various processes seems to provide
a guide for further theoretical investigations and developments.\\[1cm]

\begin{table}[h]
\caption{Investigated data.}
\begin{center}
\begin{tabular}{||l|r||l|r||}
\hline
$e^{+}e^{-}$  & $\sqrt{s}$ & $hh$  & $\sqrt{s}$ \\
\hline
TASSO $[7]$ &   22 GeV 	& $pp$ 30" B.C. at FNAL $[13]$ & 23.8 GeV \\
HRS   $[8]$ &   29 GeV 	& $pp$ SMF at CERN      $[15]$ & 30.4 GeV \\
TASSO $[7]$ & 34.8 GeV 	& $pp$ 30" B.C. at FNAL $[14]$ & 38.8 GeV \\
TASSO $[7]$	& 43.6 GeV 	& $pp$ SMF at CERN  $[15]$ & 52.6 GeV \\
ALEPH $[9]$ &   91 GeV 	& $pp$ SMF at CERN      $[15]$ & 62.2 GeV \\
DELPHI$[10]$ &   91 GeV 	& $\pbar p$ UA5 $[16]$ & 200  GeV \\
L3    $[11]$ &   91 GeV 	& $\pbar p$ UA5 $[17]$ & 546  GeV \\
OPAL  $[12]$ &   91 GeV 	& $\pbar p$ UA5 $[16]$ & 900  GeV \\
\hline
\end{tabular}
\end{center}
\end{table}

\begin{figure}
\label{DELPHI}
\vspace{500pt}
\includegraphics{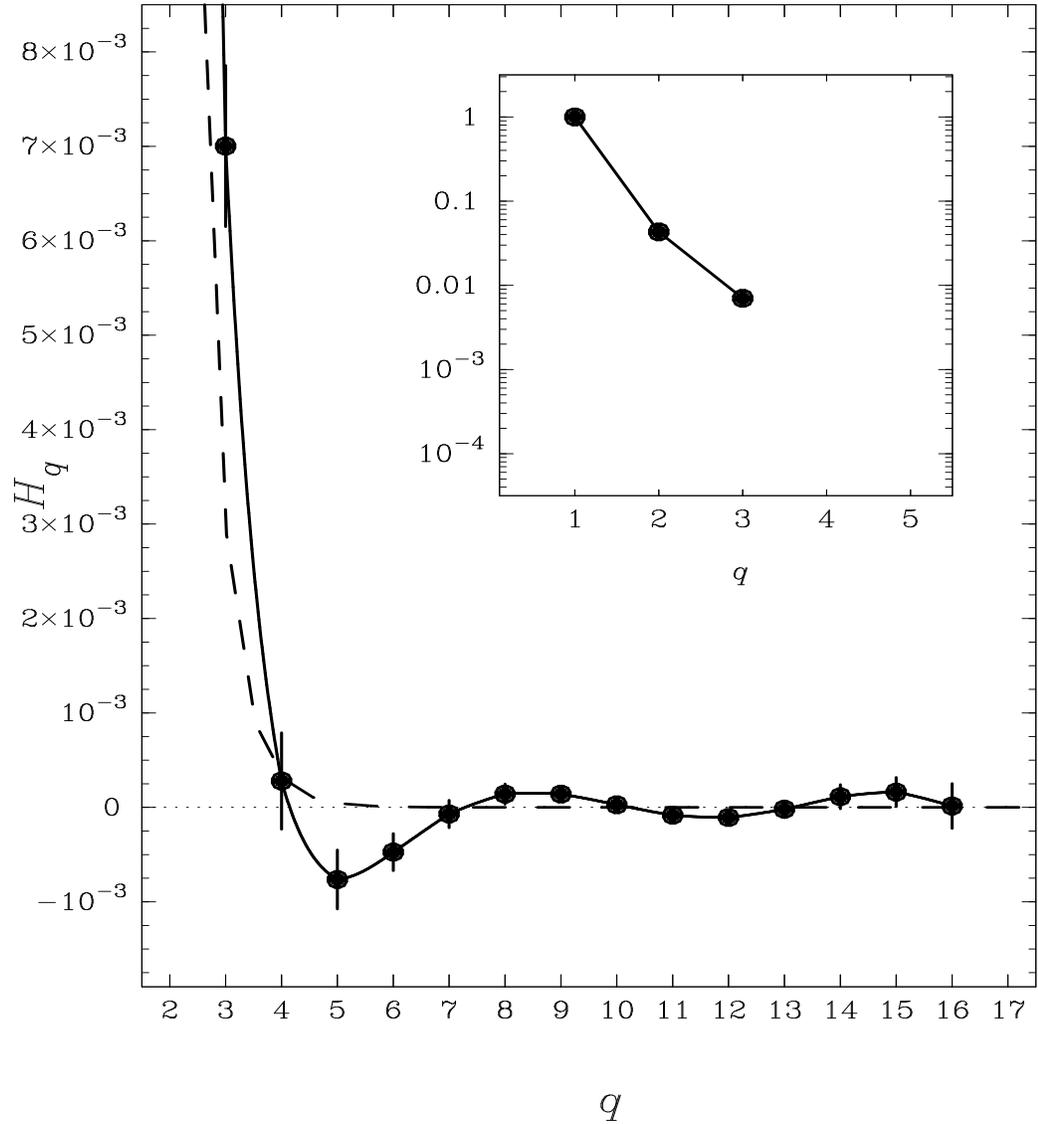}
\caption{
Example of $H_{q}$ vs. $q$ in $e^{+}e^{-}$ data,
the DELPHI experiment  (CMS energy 91 GeV).
Into the upper right box are the lower $q$ orders,
in logarithmic scale.  The markers in the main frame report the
experimental value for the $q$ orders 3 to 16; the NBD predictions
are represented by the dashed line.
}
\end{figure}
\newpage

\begin{figure}
\label{UA5}
\vspace{500pt}
\includegraphics{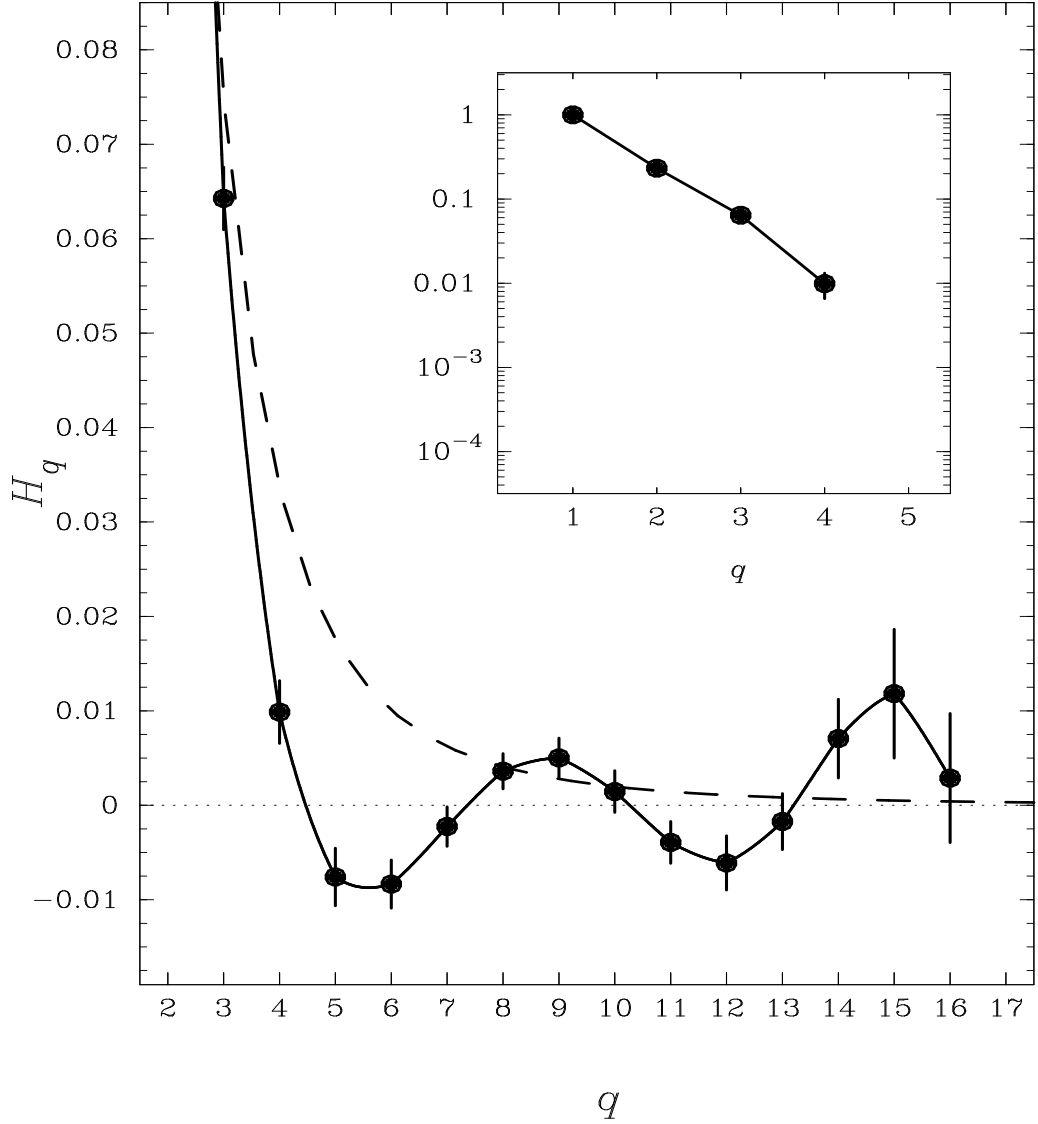}
\caption{
Example of $H_{q}$ vs. $q$ in $hh$ data,
the UA5 experiment  (CMS energy 546 GeV).
In logarithmic scale, the first four $q$ orders
are drawn in the upper box. The values from $q=3$ to $q=16$ are plotted
into the main frame. The dashed curve represents the NBD predictions.
Notice that the scale has been changed of one order of magnitude
with respect to the previous figure.
}
\end{figure}
\newpage

\newpage
\begin{figure}
\label{ALL3}
\vspace{450pt}
\includegraphics{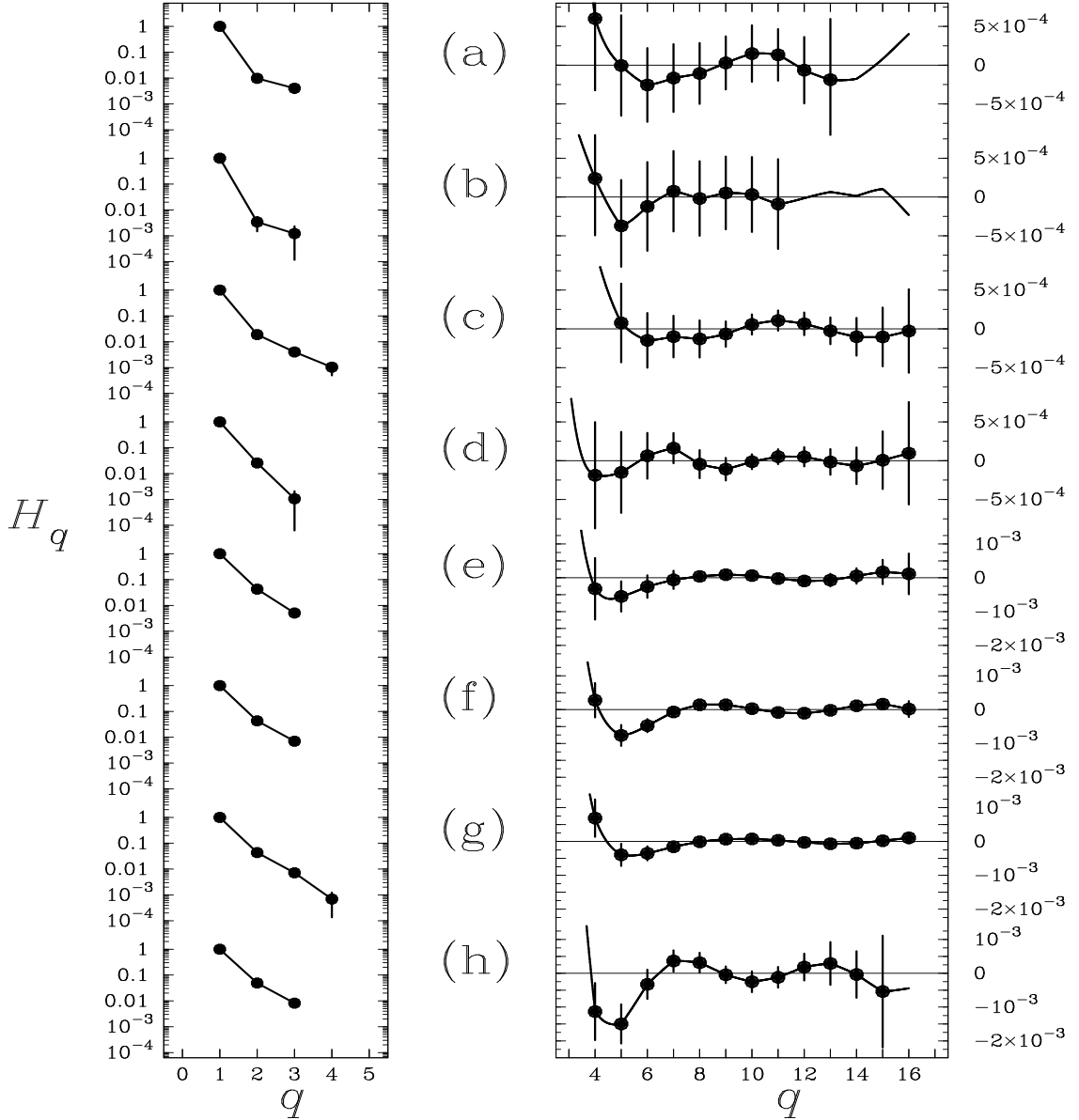}
\caption{$H_{q}$ vs. $q$ for various
$e^{+}e^{-}$ experiments. On the left the first $H_{q}$
orders in logarithmic scale, on the right the orders from
$q$=3,4 to 16.  The solid line is just meant to guide the eye
and does not represent any theoretical prediction.
The experimental outcomes are ordered
according to their energy, that increases from top to bottom:
 (a)  e+e- 22 GeV TASSO Coll.;
 (b)  e+e- 29 GeV, HRS Coll.;
 (c)  e+e- 34.8 GeV TASSO Coll.;
 (d)  e+e- 43.6 GeV TASSO Coll.;
 (e)  e+e- 91 GeV, ALEPH Coll.;
 (f)  e+e- 91 GeV DELPHI Coll.;
 (g)  e+e- 91 GeV OPAL Coll.;
 (h)  e+e- 91 GeV, L3 Coll. (here the scale has been changed
by a factor 2., with respect to the other experiments).
    }
\end{figure}

\begin{figure}
\label{ALL4}
\vspace{450pt}
\includegraphics{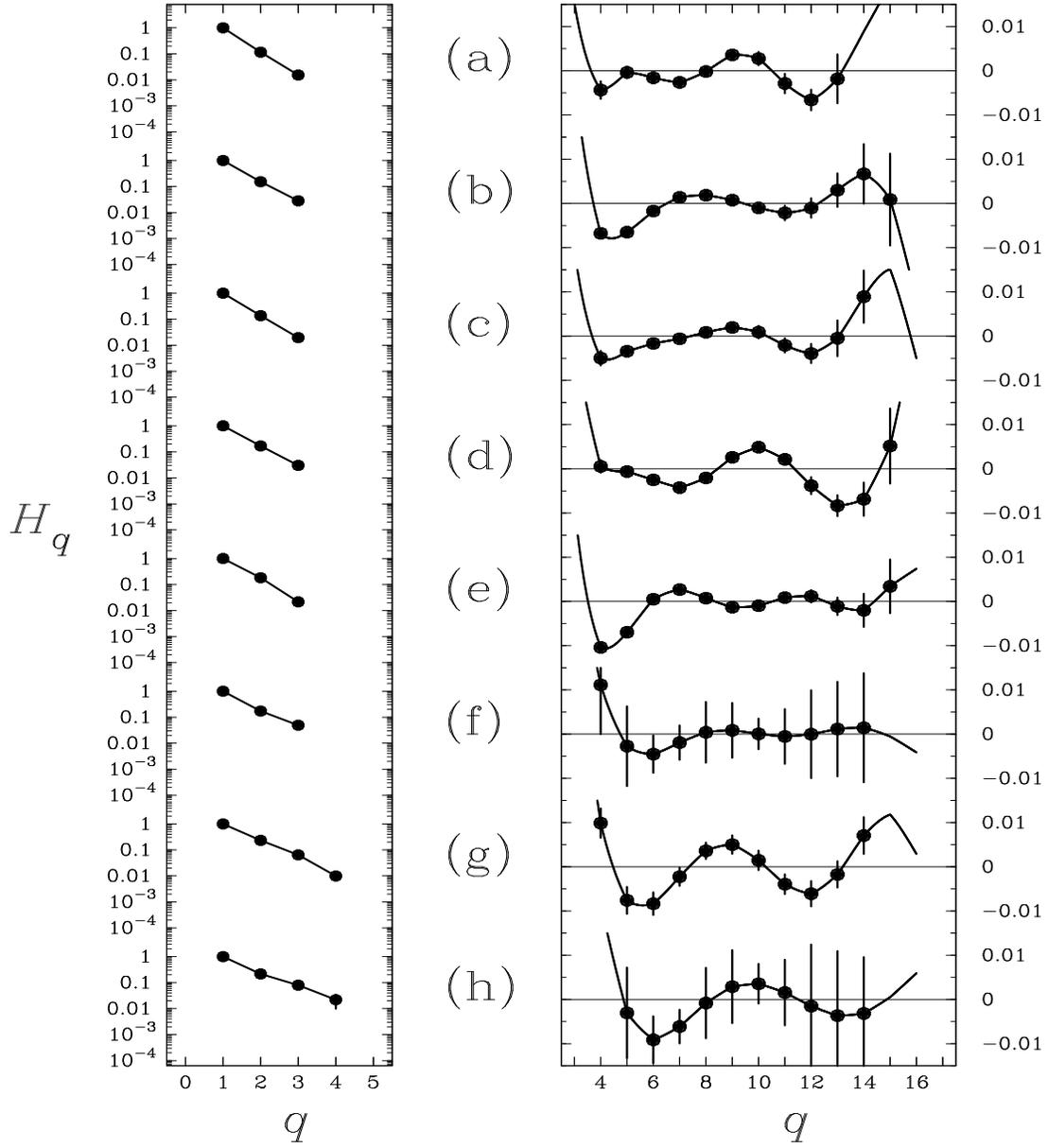}
\caption{$H_{q}$ vs. $q$ for various
$hh$ experiments. On the left the first $H_{q}$
orders in logarithmic scale, on the right the orders from
$q$=3/4 to 16 (notice that the scale on the right is different
from the one used in Fig.s 3); the solid line is just meant to guide the eye.
The points associated to exceedingly large error bars have not been plotted
for clearness and only the corresponding portion of spline has been left.
The experimental outcomes are ordered
according to their energy, that increases from top to bottom:
 (a)  pp 300 GeV/c (C.M.S. energy 23.8 GeV) FNAL 30 in. bubble chamber;
 (b)  pp 30.4 GeV SMF det. at the CERN ISR;
 (c)  pp 800 GeV/c (C.M.S. energy 38.8 GeV) E743 FNAL exp.;
 (d)  pp 52.6 GeV SMF det. at the CERN ISR;
 (e)  pp 62.2 GeV SMF det. at the CERN ISR;
 (f)  $\overline{p}p$ 200 GeV UA5 Coll.;
 (g)  $\overline{p}p$ 546 GeV UA5 Coll.;
 (h)  $\overline{p}p$ 900 GeV UA5 Coll. .
  }
\end{figure}

\end{document}